\documentclass[aps,prx,reprint,10pt,twocolumn,showpacs,superscriptaddress]{revtex4-1}
\usepackage{graphicx}
\usepackage{amssymb}
\usepackage{amsmath}
\usepackage{amsfonts}
\usepackage{color}
\usepackage[latin1]{inputenc}
\usepackage{float}
\usepackage{comment}

\newcommand{\lrangle}[1]{\langle{#1}\rangle}

\newcommand{\expct}[1]{\langle{#1}\rangle}

\begin{document}
\title{Interface fluctuations for deposition on enlarging flat substrates}
\author{I. S. S. Carrasco}
\affiliation{Departamento de F\'isica, Universidade Federal de Vi\c cosa,
36570-000, Vi\c cosa, Minas Gerais, Brazil}

\author{K. A. Takeuchi}
\affiliation{Department of Physics,\! The University of Tokyo,\! 7-3-1 Hongo,\! Bunkyo-ku,\! Tokyo 113-0033,\! Japan}%

\author{S. C. Ferreira}

\author{T. J. Oliveira}
\affiliation{Departamento de F\'isica, Universidade Federal de Vi\c cosa,
36570-000, Vi\c cosa, Minas Gerais, Brazil}

\begin{abstract}
We investigate solid-on-solid models that belong to the Kardar-Parisi-Zhang 
(KPZ) universality class on substrates that expand laterally at a constant rate 
by duplication of columns. Despite the null global curvature, we show that all 
investigated models have asymptotic height distributions and spatial covariances 
in agreement with those expected for the KPZ subclass for curved surfaces. In 
$1+1$ dimensions, the height distribution and covariance are given by the GUE 
Tracy-Widom distribution and the Airy$_2$ process, instead of the GOE and 
Airy$_1$ foreseen for flat interfaces. These results imply that, when the KPZ 
class splits into the curved and flat subclasses, as conventionally considered, 
the expanding substrate may play a role equivalent to, or perhaps more important 
than the global curvature. Moreover, the translational invariance of the 
interfaces evolving on growing domains allowed us to accurately determine, in 
$2+1$ dimensions, the analogue of the GUE Tracy-Widom distribution for height 
distribution and that of the Airy$_2$ process for spatial covariance. Temporal 
covariance is also calculated and shown to be universal in each dimension and in 
each of the two subclasses. A logarithmic correction associated to the 
duplication of column is observed and theoretically elucidated. Finally, 
crossover between regimes with fixed-size and enlarging substrates is also 
investigated. \end{abstract}


\maketitle

\section{Introduction}
\label{secintro}

The kinetic roughening of interfaces has attracted a lot of attention in the 
last decades~\cite{barabasi,meakin}. Most of the works are devoted to interfaces 
with flat asymptotic shape due to their close relation with technological 
applications as, for example, thin film growth~\cite{Evans2006,pimpinelli}. 
However, kinetic roughening with curved asymptotic shapes appears in several 
important physical systems including biological 
growth~\cite{vicsek,*bru,*galeano,*Huergo.etal-PRE2012}, topological-defect 
turbulence of nematic liquid crystals~\cite{TakeuchiPRL,*TakeuchiSP,TakeuchiJSP} 
and colloidal deposition at edges of evaporating drops~\cite{yunker}.

It is well accepted that the scaling exponents of curved and flat interfaces are 
the same within each universality 
class~\cite{krugc,masoudi1,masoudi2,Ferreira06,TakeuchiPRL,*TakeuchiSP,
TakeuchiJSP}, but the underlying fluctuations, in general, may depend on 
geometry and/or boundary
conditions~\cite{PraSpo1,[{For recent reviews on theoretical developments on the 
KPZ class, see, e.g., 
}]Kriecherbauer.Krug-JPA2010,*Corwin-RMTA2012,TakeuchiPRL,*TakeuchiSP,
TakeuchiJSP,Alves13}.
Prah\"ofer and Spohn \cite{PraSpo1} obtained an exact solution of the 
polynuclear growth (PNG) model in $d=1+1$ dimensions, in which a single seed at 
the origin as initial condition produces a macroscopically curved interface with 
fluctuations given by the Tracy-Widom (TW) distribution~\cite{tw} for the 
Gaussian unitary ensemble (GUE). Otherwise, using a line as initial condition, 
the resulting interface is macroscopically flat and the TW distribution for the 
Gaussian orthogonal ensemble (GOE) is found for underlying interface 
fluctuations.

The PNG model is known to be in the Kardar-Parisi-Zhang (KPZ) universality class, represented by the celebrated KPZ equation~\cite{kpz}
\begin{equation}
\frac{\partial h}{\partial t} = \nu \nabla^2 h + \frac{\lambda}{2} (\nabla h)^2
+ \xi,
\label{eqKPZ}
\end{equation}
where $h(x,t)$ is the height variable and $\nu$, $\lambda$ and $\xi(x,t)$ account, respectively, for the surface tension, the amplitude of nonlinear effects, and a white noise with $\expct{\xi(x,t)} = 0$ and $\expct{\xi(x,t)\xi(x',t')} = 2D\delta(x-x')\delta(t-t')$. The different height distributions for the curved and flat interfaces in the PNG model imply that the KPZ class splits at least into two subclasses, separating the curved and flat growth~\cite{PraSpo1}. Indeed, this conjecture has been confirmed recently in experiments on the topological-defect turbulence of liquid crystals~\cite{TakeuchiPRL,*TakeuchiSP,TakeuchiJSP} and in numerical simulations of models in the KPZ class~\cite{Alves11,TakeuchiJstat,Oliveira12,Alves13}. The same conclusion has also been reached analytically for a few other solvable models~\cite{Kriecherbauer.Krug-JPA2010} and in particular for the one-dimensional KPZ equation~\cite{SasaSpo1,*Amir,*Calabrese,*Imamura}.

The compilation of all results leads to the following expression, hereafter called the KPZ ansatz:
\begin{equation}
 h \simeq v_{\infty} t + s_{\lambda} (\Gamma t)^{\beta} \chi + \eta + \ldots,
\label{eqansatz}
\end{equation}
where $v_{\infty}$ and $\Gamma$ are model-dependent constant parameters, $s_{\lambda}$ is the sign of $\lambda$ in the KPZ equation~(\ref{eqKPZ}), and $\chi$ and $\eta$ are stochastic variables. The scaling exponent $\beta$ and the normalized fluctuations $\chi$ are expected to be universal. 
In particular, for $1+1$ dimensions, analytical, numerical,  and experimental studies have shown that $\chi = \chi_2 \equiv \chi_\text{GUE}$ for curved interfaces and $\chi = \chi_1 \equiv 2^{-2/3}\chi_\text{GOE}$ for flat ones, where $\chi_\text{GUE}$ and $\chi_\text{GOE}$ are the standard random variables to describe the corresponding TW distributions~\cite{PraSpo1,Kriecherbauer.Krug-JPA2010,*Corwin-RMTA2012,TakeuchiPRL,*TakeuchiSP,TakeuchiJSP,Alves13}. The applicability of the ansatz (\ref{eqansatz}) to $2+1$ dimensions, with distinct universal distributions for flat and curved growth, was recently reported~\cite{healy12,healy13,Oliveira13} and experimentally verified, for the flat case, in the growth of semiconductor~\cite{almeida2013} and organic~\cite{healy2014} films. Furthermore, equation~(\ref{eqansatz}) was numerically shown to hold for the restricted solid-on-solid (RSOS) model on dimensions $d$ at least up to $d=6+1$~\cite{Alves14}.

Evolving curved interfaces investigated up to now are hallmarked by both 
macroscopic curvatures and expanding activity domains, whereas in flat growth 
this domain size (the substrate size) is kept constant. Therefore, a basic 
question arises: Is the curvature responsible for the appearance of the 
different distributions in the KPZ class, or whether the growth domain expands 
or not drives the height fluctuations to the different universal distributions? 
In order to address this question, we study standard flat-interface models in 
the KPZ class on substrates whose lateral size increases at a constant rate 
$\omega$ but the macroscopic curvature is kept null. Scaling exponents for 
interface growth models on expanding domains were recently 
analyzed~\cite{Pastor,Escudero2,masoudi1}. Since the spatial correlation length 
increases as $\xi_{\parallel} \sim t^{1/z}$, where $z$ is the dynamic exponent, 
for a substrate increasing as $L\sim t^{\gamma}$, the interface width evolves 
indefinitely as $W \sim t^{\beta}$ if $\gamma=1>1/z$, because  correlation 
length never reaches the system size~\cite{Pastor}. Otherwise, for $\gamma<1/z$  
the surface becomes completely correlated $(\xi \sim L)$ after a crossover time 
and the interface width scales as $W \sim t^{\gamma \alpha}$~\cite{Pastor}, 
where $\alpha=\beta z$ is the roughness exponent. Similar behavior was found in 
an analytical study of linear growth equations on growing 
domains~\cite{Escudero2}. Masoudi \textit{et al}.~\cite{masoudi1} analyzed some 
typical flat models on substrates which grow at a constant rate ($\gamma=1$), by 
alternating deposition and substrate enlargement deterministically, and obtained 
the same growth exponents as for the fixed-size case.

In the present work, the substrate enlargement is performed stochastically, by 
duplicating randomly selected columns at a rate $\omega$ in addition to the 
usual deposition rules. We show that expanding systems exhibit height 
distributions given by the GUE TW distribution in $d=1+1$ and its counterpart in 
$d=2+1$, showing that they belong to the same KPZ subclass as the curved 
interfaces. This is also confirmed by the spatial covariance, given by the 
Airy$_1$ and Airy$_2$ process for the fixed and growing domains, respectively, 
in $d=1+1$, and their counterparts in $d=2+1$. Universality in temporal 
covariance is also shown in $d=1+1$ and $2+1$, again, with different universal 
functions for the different KPZ subclasses. The duplication mechanism introduces 
logarithmic corrections in the KPZ ansatz, which are explained with an 
approximate theoretical analysis. Furthermore, analyzing the effects of the 
initial size of the substrate, we characterize crossover from the fixed-size 
(GOE in $1+1$) to the enlarging substrate (GUE in $1+1$) regimes.

This paper is organized as follows. In Sec.~\ref{secmodels} we define the 
studied models and the method of substrate expansion. 
Sections~\ref{secresults1d} and \ref{secresults2d} present the height 
distribution analysis for one and two-dimensional substrates, respectively. The 
spatial and temporal covariances are presented in Secs.~\ref{seccovspatial} 
and~\ref{seccovtemp},  respectively, and the crossover effect controlled by the 
initial substrate size in Sec. \ref{seccross}. Section \ref{secconcl} summarizes 
our conclusions and final discussions.

\section{Growth models on enlarging domains}
\label{secmodels}

We study the restricted solid-on-solid (RSOS)~\cite{kk}, the single step 
(SS)~\cite{barabasi} and the Etching~\cite{Mello} models on enlarging substrates 
represented by chains in $d=1+1$ and square lattices in $d=2+1$, with periodic 
boundary conditions. In all models, particles are added at a randomly chosen 
site $i$ according to the following rules: RSOS - if $h_{j}-h_{i} = 0$ or 1 for 
$\forall j\in\mathcal{N}(i)$, then $h_i \rightarrow h_i + 1$ (so that 
$|h_{j}-h_{i}| \leq 1$ is always satisfied); SS - if $h_{j}-h_{i} = 1$ for 
$\forall j\in\mathcal{N}(i)$, then $h_i \rightarrow h_i + 2$; Etching - $h_i 
\rightarrow h_i + 1$ and, if $h_{j} < h_{i}-1$, then $h_{j}\rightarrow  
h_{i}-1$ for each $j\in\mathcal{N}(i)$. Here, $\mathcal{N}(i)$ represents the 
set of the nearest neighbors (NN) of $i$. Flat initial conditions, $h_i=0$, were 
used for RSOS and Etching models while chessboard initial conditions,  $h_{i}$ 
alternating between 0 and 1, were used for the SS model.

\begin{figure}[!b]
\includegraphics[width=8cm]{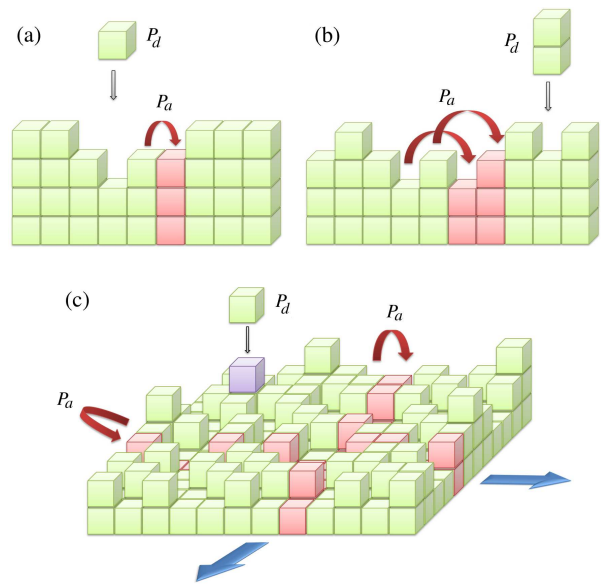}
\caption{(Color online) Illustration of the substrate enlargement and deposition
for (a) RSOS and (b) SS models in $1+1$ and in (c) $2+1$ dimensions.}
\label{fig1}
\end{figure}

The substrate enlargement is implemented as follows. A particle deposition is 
attempted with probability $P_d = N/(N+\omega d_s)$ while a column duplication 
occurs with complementary probability $P_a = \omega d_s/(N+\omega d_s)$, where 
$N$ is the number of the lattice sites and $d_s$ is the substrate dimension. 
After each event, time is increased by $\Delta t=1/(N+\omega d_s)$. The initial 
lateral substrate size is $L_0$. In $d=1+1$, each substrate enlargement is 
implemented by a simple, local duplication of a randomly selected column for the 
RSOS and Etching models, as illustrated in Fig.~\ref{fig1}(a). In $d=2+1$, a 
lattice row or column is randomly selected and similarly duplicated, as 
illustrated in Fig.~\ref{fig1}(c). The lateral lattice size increases on 
average, therefore, as $\lrangle{L}=L_0+\omega t$. In the SS model, we must 
duplicate a pair of NN columns at once to conserve the steps at surface, see 
Fig.~\ref{fig1}(b). Therefore, the substrate enlarging rate is $2\omega$.

Let us illustrate a consequence of the column duplication for the KPZ ansatz by an approximate argument. Let $\nabla h_i$ be the local gradient on a 
$d_s$-dimensional substrate, so that, $(\nabla h_i)^2 = \left(\frac{\partial 
h_i}{\partial x_1} \right)^2 + \left(\frac{\partial h_i}{\partial x_2} \right)^2 
+ \cdots + \left(\frac{\partial h_i}{\partial x_{d_s}} \right)^2$, where $x_j$ 
with $j=1,\cdots,d_s$ are the substrate directions. The mean squared gradient
at time $t$ is \[ G_{t} = \frac{1}{L^{d_s}} \sum_{i=1}^{L^{d_s}} (\nabla h_i)^{2}.\]
After $\omega$ duplications, in a time unity, we have
\begin{equation}
G_{t+1} = \frac{1}{(L+\omega)^{d_s}} \left[ \sum_{i=1}^{L^{d_s}} (\nabla h_i)^2 + \sum_{i=1}^{(L+\omega)^{d_s}-L^{d_s}} (\nabla h_i)^2 \right],
\label{eqGrad}
\end{equation}
where the first and second sums are taken over non-duplicated and duplicated 
sites, respectively. Considering only the effects of duplication and using the 
statistical equivalence of sites, we have 
\[\sum_{i=1}^{L^{d_s}} (\nabla h_i)^2 \simeq L^{d_s} G_t \] 
and 
\[\sum_{i=1}^{(L+\omega)^{d_s}-L^{d_s}} (\nabla h_i)^2 \approx \frac{(d_s-1)}{d_s} \left[(L + \omega)^{d_s} -L^{d_s}\right] G_t, \] 
where the ratio $(d_s - 1)/d_s$ appears because column duplication in direction $x_k$ implies $\left(\frac{\partial h_i}{\partial x_k} \right)^2 = 0$ along this column, right after the duplication. Inserting this result in Eq.~(\ref{eqGrad}) and considering long times, so that $L \sim \omega t$, we find \[G_{t+1} \approx \left\lbrace  1 + \frac{1}{d_s} \left[-1  + \frac{1}{(1 + 1/t)^{d_s}} \right] \right\rbrace  G_t.\]
Therefore, disregarding terms $\mathcal{O}(t^{-2})$ for $t \gg 1$, we have \[G_{t+1} - G_{t} \approx -\frac{1}{t} G_t \quad \quad \text{or} \quad \quad \frac{d G}{dt} \approx
-\frac{1}{t} G,\] implying  $G_t \sim 1/t$ due to the substrate expansion. 
It is unclear whether the column duplication produces the same effect when the particle deposition is also considered, but the simplest scenario would be to assume that the above functional form of $G_t$ describes an additive correction to the height evolution~(\ref{eqKPZ}), induced by the column duplication. This implies the presence of a logarithmic correction to 
the KPZ ansatz [Eq.~(\ref{eqansatz})], which now reads
\begin{equation}
 h \simeq v_{\infty} t + s_{\lambda} (\Gamma t)^{\beta} \chi + \eta + s_{\lambda} \zeta \ln t +
\ldots
\label{eqansatz2}
\end{equation}
where $\zeta$ is, in principle, a stochastic variable. We will see that this logarithmic correction indeed exists in all models and dimensions we investigated, and that the fluctuations of $\zeta$, if exist, are very small.
Note however that this logarithmic correction is predicted for the KPZ-class 
interfaces on expanding substrates and not necessarily for other universality 
classes. More importantly, one can easily see that the duplication of a column does not induce any curvature in the global scale, i.e., \[\lrangle{\nabla^2 h} = 0,\] 
which is guaranteed here by the choice of the periodic boundary condition. 
This allows us to study the effect of the substrate expansion on the KPZ universal fluctuations, independently of the global curvature.

\section{Height fluctuations in $1+1$ dimensions}
\label{secresults1d}

\begin{figure}[!t]
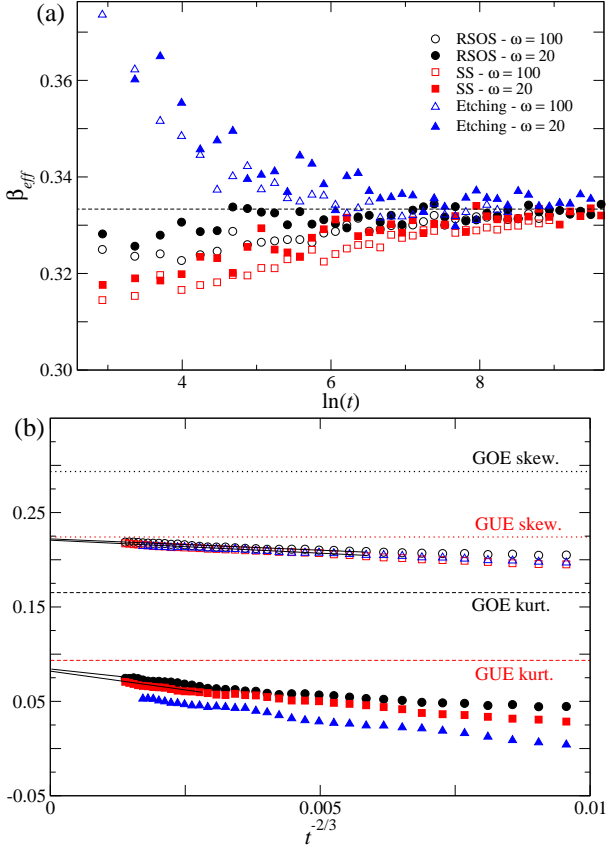

\includegraphics[width=8cm]{Fig2a.eps}
\includegraphics[width=8cm]{Fig2b.eps}
\caption{(Color online) (a) Effective growth exponent as a function of $\ln(t)$. 
(b) Skewness (open) and kurtosis (full symbols) of the height distributions for 
the RSOS (black circles), SS (red squares) and Etching (blue triangles) models, 
with $\omega=20$. Dashed and dotted horizontal lines indicate the expected 
values of $S$ and $K$ for the GUE and GOE TW distributions, respectively.}
\label{fig2}
\end{figure}

This section presents numerical results for the one-dimensional enlarging 
substrates, with $L_{0}=\omega$ and up to $25000$ realizations. Figure 
\ref{fig2}(a) shows the effective growth exponent, defined by $\beta_{\rm 
eff}(t) \equiv \frac{1}{2}\frac{d(\log \expct{h^2}_c)}{d(\log t)}$ with the second-order cumulant $\expct{h^2}_c$,
for all investigated models and two different values 
of $\omega$. The convergence to the expected KPZ value $\beta=1/3$ is found in 
all cases, in agreement with previous simulations of KPZ models in linearly 
growing domains, $L\sim t$~\cite{Pastor,masoudi1}.

To characterize  the asymptotic height distributions, we analyzed the 
dimensionless cumulant ratios $S=\left\langle h^{3} \right\rangle_c / 
\left\langle h^{2} \right\rangle_c^{3/2}$ (skewness) and $K=\left\langle h^{4} 
\right\rangle_c / \left\langle h^{2} \right\rangle_c^{2}$ (kurtosis), where
$\left\langle h^{n} \right\rangle_c$ is the $n$th-order cumulant of $h$. The results are plotted in Fig.~\ref{fig2}(b) as functions of $t^{-2\beta}$, which is an expected functional form for their finite-time correction, usually obtained on the basis of the correction of $\mathcal{O}(t^{-2\beta})$ in the second-order cumulant \cite{Oliveira12,Alves13,TakeuchiJSP,Ferrari.Frings-JSP2011}.
Our results indeed underpin this finite-time correction, and, 
extrapolating the data, we find that the asymptotic skewness and kurtosis 
indicate the values for the GUE TW distribution. We conclude, therefore, that 
the underlying distribution behind the asymptotic height fluctuations is given 
by the GUE TW distribution, rather than the GOE counterpart found for 
$\omega=0$. It is important to emphasize that the global curvature of the 
interface is not changed by duplications  and remains identically null due to 
the periodic boundary conditions.

As $t \rightarrow \infty$, the number of the sites duplicated during each time 
unit becomes negligible compared with that of the non-duplicated sites. 
Therefore, the non-universal parameters such as $v_{\infty}$, $\lambda$ and 
$\Gamma$ should not be changed by the substrate expansion. 
This implies that it is sufficient for us to determine them for the non-expanding case, $\omega = 0$.
For the RSOS model, $v_{\infty} = 0.419030(3)$ and $\Gamma = 0.252(1)$ were numerically estimated in 
\cite{Oliveira12,Alves13}. The exact values of these quantities for the SS model 
are $v_{\infty} = \Gamma = 1/2$ \cite{KrugPRA92}. Following the same procedures 
as in Refs.~\cite{Oliveira12,TakeuchiJstat}, we found $v_{\infty} = 
2.13986(5)$ and $\Gamma = 4.90(9)$ for the Etching model. All these results were 
obtained for $\omega=0$, but the validity of these values for the expanding case was 
explicitly checked.

\begin{figure}[!t]
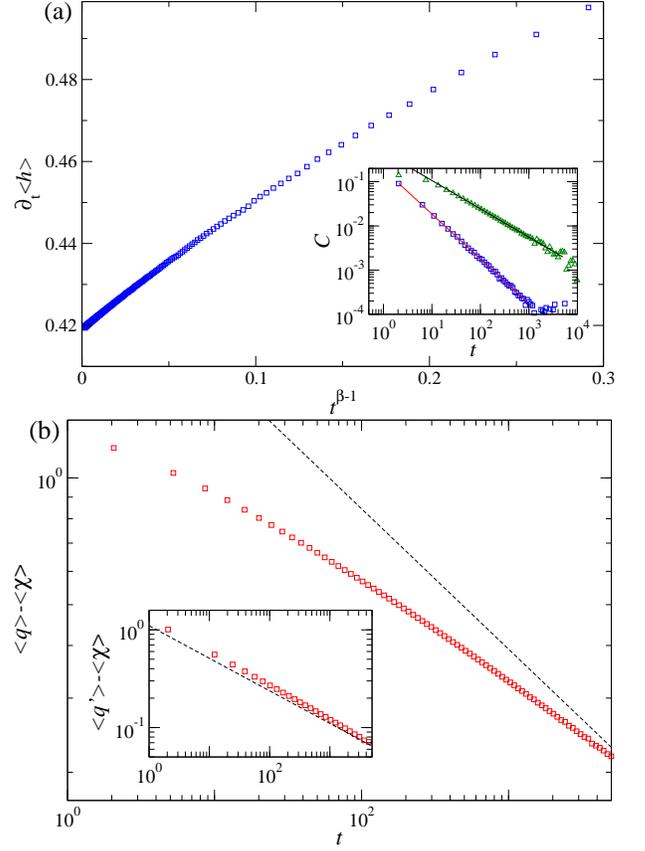

\includegraphics[width=8cm]{Fig3a.eps}
\includegraphics[width=8cm]{Fig3b.eps}
\caption{(Color online) (a) Growth velocity versus $t^{\beta-1}$. Inset shows
$C \equiv \partial_t \left\langle h\right\rangle - v_{\infty} - s_{\lambda} \beta \Gamma^{\beta} \left\langle \chi \right\rangle t^{\beta-1}$ against $t$ with $\left\langle \chi \right\rangle = \left\langle \chi_1 \right\rangle$ (green triangles) and $\left\langle \chi \right\rangle = \left\langle \chi_2 \right\rangle$ (blue squares). The lines are results of power-law regressions. (b) Finite-time correction in the mean height versus time, disregarding (main plot) or considering (inset) the logarithmic correction (see text). Dashed lines indicate the slope $-1/3$. All data above were obtained for the RSOS model with $\omega=20$. 
}
\label{fig3}
\end{figure}

Accordingly to Eq.~(\ref{eqansatz}), we have
\[\partial_t \left\langle h\right\rangle \simeq v_{\infty} +
s_{\lambda} \beta \Gamma^{\beta} \left\langle \chi
\right\rangle t^{\beta-1}.\]
Then, plotting $\partial_t \left\langle h\right\rangle$ against $t^{\beta-1}$ 
should result in a straight line, whose $y$-intercept is $v_{\infty}$. However, 
we did not find such linear behavior even for the longest times investigated, as 
shown in Fig.~\ref{fig3}(a). Indeed, the additional logarithmic correction 
predicted in Eq.~(\ref{eqansatz2}) can not be neglected.
Assuming that 
\begin{equation}
\partial_t \left\langle h\right\rangle \simeq v_{\infty} + s_{\lambda} \beta
\Gamma^{\beta} \left\langle \chi \right\rangle t^{\beta-1} + s_{\lambda} \expct{\zeta} t^{-\delta}
+ \ldots,
\label{eqvelocity}
\end{equation}
the correction $s_{\lambda} \expct{\zeta} t^{-\delta}$ can be obtained by plotting $C \equiv \partial_t
\left\langle h\right\rangle - v_{\infty} - s_{\lambda} \beta \Gamma^{\beta} 
\left\langle \chi \right\rangle t^{\beta-1}$ against time, as shown in the inset 
of Fig.~\ref{fig3}(a). For all investigated models, we found the exponent 
$\delta = 1.01(1)$, consistent with the logarithmic correction, using 
$\left\langle \chi \right\rangle = \lrangle{\chi_2}$ (GUE TW). Instead, if the 
GOE TW value is used, $\delta \approx 1-\beta = 2/3$ is found (see inset of 
Fig.~\ref{fig3}(a)). This indicates that the term $t^{\beta-1}$ in 
Eq.~(\ref{eqvelocity}) was not absorbed if the GOE TW value is assumed, giving 
further evidence that the GOE TW distribution does not correctly describe the 
distribution of $\chi$.

The importance of the logarithmic term in Eq.~(\ref{eqansatz2}) is evidenced when we try to determine the usual finite-time correction term $\eta$. Defining the variable \[q \equiv \frac{h-v_{\infty}t}{s_{\lambda}(\Gamma t)^{\beta}},\] the finite-time correction in the mean height has a power law decay, $\left\langle q\right\rangle - \left\langle \chi \right\rangle\sim t^{-\beta}$,  for $\omega=0$~\cite{Oliveira13,Alves13,TakeuchiPRL,*TakeuchiSP,TakeuchiJSP,Alves11,TakeuchiJstat,Oliveira12}. However, for $\omega>0$, a power law decay is not found due to the logarithmic correction,
as shown in Fig.~\ref{fig3}(b). Instead, including the logarithm term as
\begin{equation}
q' \equiv \frac{h-v_{\infty}t - s_{\lambda} \expct{\zeta} \ln t}{s_{\lambda} (\Gamma
t)^{\beta}},
\label{eqqprime}
\end{equation}
and using the value of $\expct{\zeta}$ estimated from the power law regressions shown in the  inset of 
Fig.\ref{fig3}(a), the finite-time correction decaying as $t^{-\beta}$ is 
recovered (see inset of Fig. \ref{fig3}(b)). Moreover, we observe that 
$\expct{\zeta}$ does not significantly depend on $\omega$ and take values 
$\expct{\zeta}=0.18(1)$, $0.15(4)$ and $0.22(3)$ for the RSOS, SS and Etching 
models, respectively, for $\omega$ varying from $1$ to $100$. 
Our analysis does not permit a conclusive assessment on
logarithmic corrections in higher-order cumulants of the height distribution. 
For SS model, for which $\Gamma=1/2$ is exactly known, 
$\expct{h^n}_c - [s_\lambda(\Gamma t)^\beta]^n\expct{\chi^n}_c$
 seems to reach a constant value, suggesting that $\zeta$ is deterministic.
For RSOS, a very small variance $\lrangle{\zeta^2}_c \approx 0.005$, 
smaller than the uncertainties obtained with our current precision in $\Gamma$,
 is found.


\begin{figure}[!t]
\includegraphics[width=8.5cm]{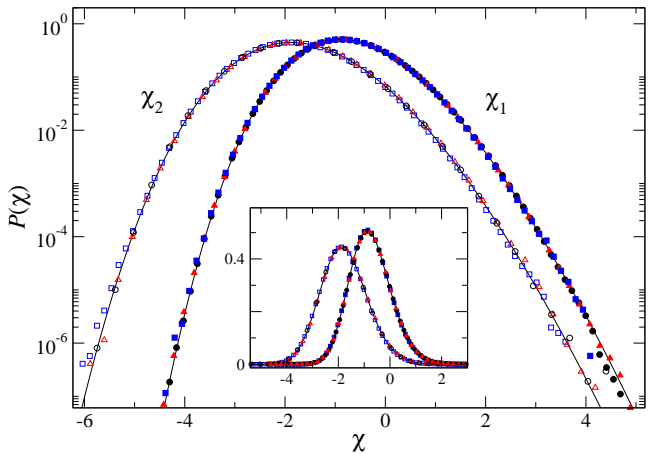}
\caption{(Color online) Rescaled height distributions for the RSOS (black 
circles), SS (red triangles) and Etching (blue squares) models, with $\omega = 
20$ (open symbols) for times $t=18000$, 19000 and 15000, respectively, and with 
$\omega = 0$ (full symbols) for $t=2000$, 2000 and 1500, respectively. Main plot 
and inset show these data in semi-log and linear scales, respectively. Here, 
histograms of $\chi \equiv (h-v_{\infty}t - \expct{\eta} - s_{\lambda} \zeta \ln 
t)/(\Gamma t)^{\beta}$ are compared with the probability density of $\chi_1$ and 
$\chi_2$.}
\label{fig4}
\end{figure}

Similarly to $\expct{\zeta}$,
$\lrangle{\eta}$ also seems to be independent of $\omega$: we found
$\lrangle{\eta} = -0.87(3)$, -0.44(3) and 3.2(2) for the RSOS, SS and Etching 
models, respectively, for the values of $\omega$ we investigated ($\omega \geq 
1$). However, they are larger (in the absolute value) than the estimates for 
$\omega=0$, which are $\lrangle{\eta}=-0.32(4)$ for the RSOS 
model~\cite{Alves13}, and $\lrangle{\eta}=-0.33(1)$ for the SS and 
$\lrangle{\eta}=0.20(3)$ for the Etching models (present work). Note that, for 
short times,  the surface roughness is very small and duplication of columns 
does not have significant effects on the mean height, so that $\left\langle h 
\right\rangle_{\omega>0} \approx \left\langle h \right\rangle_{\omega=0}$ for 
short times. This suggests that $\lrangle{\eta}$ becomes larger for $\omega>0$ 
so as to compensate the change in the mean height due to the crossover from 
the GOE TW distribution to the 
GUE counterpart (see Sec.~\ref{seccross}).

Height distributions rescaled according to Eq.~(\ref{eqansatz2}) are presented 
in Fig.~\ref{fig4}. Excellent data collapse with the theoretical curve for the 
GUE TW distribution demonstrates
that the asymptotic height fluctuations of flat models on the growing substrates 
are given by the GUE TW distribution. The distributions for the same models with 
$\omega=0$ are also shown for the sake of comparison. It is important to remark 
that rescaled height distributions without taking into account the logarithmic 
correction display a shift in the mean decaying very slowly as 
$\ln(t)/t^{\beta}$.

\section{Height fluctuations in $2+1$ dimensions}
\label{secresults2d}

For two-dimensional enlarging substrates, we used $L_0 = \omega$ in both 
directions, with $\omega$ varied from $1$ to $10$, and averages were taken over 
$40000$ samples.

\begin{figure}[!t]
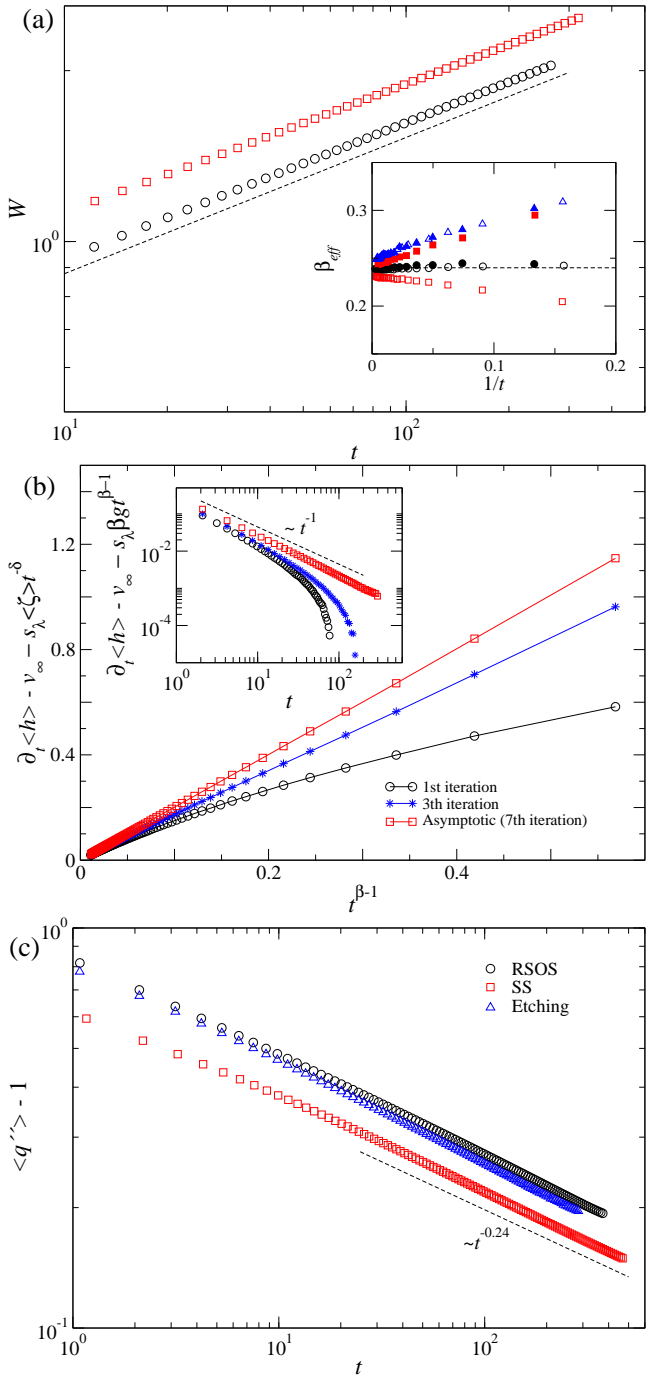

\includegraphics[width=8.5cm]{Fig5a.eps}
\includegraphics[width=8.5cm]{Fig5b.eps}
\includegraphics[width=8.5cm]{Fig5c.eps}
\caption{(Color online) (a) Interface width versus time for the RSOS (black 
circles) and SS (red squares) models, with $\omega = 10$. Inset shows the 
effective growth exponents for the RSOS, SS and Etching (blue triangles) models, 
with $\omega=6$ (full) and $10$ (open symbols). Dashed line has the slope 
$\beta=0.24$. (b) Illustration of the method to self-consistently determine $g 
\equiv \Gamma^{\beta} \expct{\chi}$ (main plot) and $\expct{\zeta}t^{-\delta}$ 
(inset) for the RSOS model with $\omega=6$. We start the first iteration with 
$\expct{\zeta}t^{-\delta} = 0$. (c) Shift in rescaled mean height versus time 
for the three models with $\omega=6$.}
\label{fig5}
\end{figure}

Figure~\ref{fig5}(a) shows the interface width ($W \equiv \sqrt{\left\langle 
h^{2}\right\rangle_c}$) evolution in time and the corresponding effective growth 
exponent. Analogously to the one-dimensional case, the growth exponent values 
converge to that of the KPZ  class in $d = 2+1$ dimensions, $\beta \approx 
0.24$~\cite{Kelling}. The asymptotic growth velocities $v_{\infty}$ for the 
three models with $\omega=0$ were calculated in~\cite{Oliveira13} (see Table I). 
For $\omega>0$, we find behavior of $\partial_t \left\langle h\right\rangle$ 
analogous to that for $d=1+1$ (Fig.~\ref{fig3}(a)), but since here the exact 
value of $\expct{\chi}$ is not known, we determine $\expct{\zeta}$ and $\delta$ 
in Eq.~(\ref{eqvelocity}) in an iterative way as follows. Plotting $\partial_t 
\left\langle h\right\rangle - v_{\infty}$ against $t^{\beta-1}$, we estimate a 
rough value for the product $g \equiv \Gamma^{\beta} \expct{\chi}$ from the 
slope near the origin (Fig. \ref{fig5}(b)). We then use this 
\textit{approximate} value to plot $\partial_t \left\langle h\right\rangle - 
v_{\infty} - s_{\lambda} \beta g t^{\beta-1}$ against $t$ in logarithmic scales 
(inset of Fig. \ref{fig5}(b)) as done in Fig.~\ref{fig3}(a). We expect it to 
decay as $\expct{\zeta} t^{-\delta}$ with $\delta > 1-\beta$, but because of the 
error in the estimate of $g$, the residual in the order of $t^{\beta-1}$ 
dominates for large $t$. Therefore, $\expct{\zeta}$ and $\delta$ can be 
estimated from the data at small $t$. With these estimates, we reestimate $g$ 
by plotting $\partial_t \left\langle h\right\rangle - v_{\infty} - 
s_\lambda\expct{\zeta}t^{-\delta}$ against $t^{\beta-1}$ (see again the main 
panel of Fig. \ref{fig5}(b)). Repeating this procedure to improve the estimates 
$g, \expct{\zeta}, \delta$ until they reach some asymptotic values, we finally 
find straight lines in both plots (red squares), which guarantee the 
self-consistency of the estimates. In particular, we find $\delta \approx 1$ 
which indicates the presence of a logarithmic correction in the KPZ ansatz also 
for $d=2+1$, as expected by the theoretical argument presented in 
Sec.~\ref{secmodels}. As in $d=1+1$, $\lrangle{\zeta}$ is almost independent of 
$\omega$ within the range of $\omega$ studied here, taking values at 
$\lrangle{\zeta}=0.32(2), 0.36(4), 0.39(2)$ for the RSOS, SS and Etching models, 
respectively.

To determine $\expct{\eta}$, following Refs. \cite{Alves13,Oliveira13},
 we define the variable
\begin{equation}
 q'' \equiv \frac{h-v_{\infty}t - s_{\lambda} \expct{\zeta} \ln t}{s_{\lambda} g t^{\beta}},
\end{equation}
so that $\expct{q''} - 1 \simeq (s_{\lambda} \expct{\eta}/g)t^{-\beta}$. Figure 
\ref{fig5}(c) shows this shift against time, where the expected decay 
$t^{-\beta}$ is observed and, using the values of $g$ obtained above, we 
estimate $\expct{\eta}$ at $\left\langle \eta \right\rangle = -1.8(2)$, -1.4(1), 
and 4.9(1) for the RSOS, SS and Etching models, respectively. These values are 
again independent of $\omega$ within $1 \leq \omega \leq 10$ and larger than the 
values for $\omega=0$ (see Table I in Ref.~\cite{Oliveira13}).

\begin{figure}[!t]
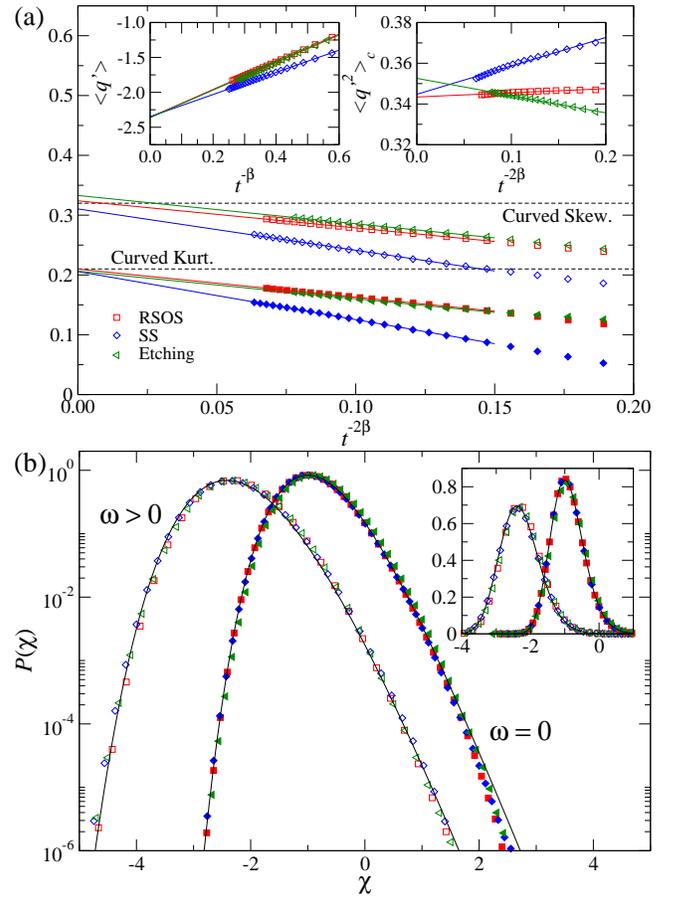

\includegraphics[width=8.5cm]{Fig6a.eps}
\includegraphics[width=8.5cm]{Fig6b.eps}
\caption{(Color online) (a) Skewness (open) and kurtosis (full symbols) against 
$t^{-2\beta}$ for all investigated models with $\omega=10$. Insets show 
$\expct{q'}$ (left) and $\expct{q'^{2}}_c$ (right) versus $t^{-\beta}$ and 
$t^{-2\beta}$, respectively, for the same models. (b) Rescaled height 
distributions for RSOS, SS and Etching models, with $\omega = 4$ (open symbols) 
and times $t=500$, 650 and 350, respectively, and, with $\omega = 0$ (full 
symbols) and $t=10000$, 8000 and 2000, respectively. Main panel and inset show 
these data in semi-log and linear scales, respectively. Full lines are 
generalized Gumbel distributions with $m=9.5$ (left) and $m=6.0$ 
(right)~\cite{Oliveira13}.}
\label{fig6}
\end{figure}

The parameter $\Gamma$ is given by $\Gamma = (1/2)|\lambda|A^{1/\alpha} = 
|\lambda|A^2/2$ for $d=1+1$ dimensions and $\Gamma = |\lambda|A^{1/\alpha}$ for 
$d=2+1$ dimensions, where $A$ is defined by the steady-state growth velocity 
$v_s(L)$ in a system of (fixed) size $L$, through $v_s(L) - v_s(\infty) = - 
(\lambda A/2)L^{2\alpha-2}$ \cite{Krug90,KrugPRA92,healy2014}. Note that the 
presence or absence of the factor 1/2 in the above expressions for $\Gamma$ is 
not essential, but introduced only to conform with the definitions adopted in 
past studies. The parameter $\lambda$ can be obtained from the dependence of the 
asymptotic growth velocity on the substrate slope $u$ \cite{Krug90}; 
specifically, $\lambda = \left( \frac{\partial^{2} v_{\infty}}{\partial u^2} 
\right)_{u\rightarrow 0}$. Therefore, by plotting $v_s(L) - v_s(\infty)$ against 
$L^{2\alpha-2}$ with the value $\alpha = 0.39$ for the (2+1)-dimensional KPZ 
class \cite{marinari,Kelling}, and by using all the expressions above, we 
determined $A$ and $\Gamma$ as listed in Table I.

\begin{table*}[bth]
\begin{center}
\begin{tabular}{c c c c c c c c c c c c c c c c c}
\hline \hline
& & $v_\infty$ \cite{Oliveira13} & & $\lambda$ & & $A$ & & $\Gamma$ & &
$\langle\chi\rangle_c$ & & $\langle\chi^2\rangle_c$ & & $S$ & & $K$ \\
\hline
RSOS       & & 0.31270(1)  & & -0.405(7) & & 1.22(4)  & & 0.68(6)  & & -2.34(3)
& & 0.341(5) & & 0.328(4) & & 0.210(4) \\
SS         & & 0.341368(3) & & -0.481(3) & & 1.44(5)  & & 1.2(1)   & & -2.37(5)
& & 0.336(6) & & 0.329(7) & & 0.206(3) \\
Etching    & & 3.3340(1)   & & 2.147(4)  & & 3.629(9) & & 58.5(5)  & & -2.36(3)
& & 0.346(8) & & 0.336(6) & & 0.21(1)  \\
\hline\hline
\end{tabular}
\caption{Non-universal ($v_{\infty}$, $\lambda$, $A$ and $\Gamma$) and universal
($\langle\chi\rangle_c$, $\langle\chi^2\rangle_c$, $S$ and $K$) quantities for
different models on enlarging  $d=2+1$ dimensional substrates. The averages and uncertainties of universal
quantities were determined using different substrate expansion rates $\omega>0$.}
\label{tab1}
\end{center}
\end{table*}

Using the estimated parameter values to rescale the height as in 
Eq.~(\ref{eqqprime}), universality in the height distribution for $d=2+1$ can be 
explicitly assessed. The mean and the variance of $\chi$ are obtained from 
extrapolations of $\left\langle q'\right\rangle $ and $\left\langle 
q'^{2}\right\rangle_c$ against $t^{-\beta}$ and $t^{-2\beta}$, respectively, as 
shown in the insets of Fig. \ref{fig6}(a). Table I summarizes the values 
obtained by using different substrate expansion rates $\omega>0$. These values 
are in good agreement with those obtained by Halpin-Healy \cite{healy12,healy13} 
for curved interfaces and far from those for the flat ones. Therefore, the 
underlying fluctuations in two-dimensional enlarging substrates are also 
equivalent to those found for curved systems. This is corroborated by the 
skewness and the kurtosis of the height distributions, which converge to the 
corresponding values for the curved interfaces, as shown in Fig. \ref{fig6}(a) 
and Table I. It is worth stressing again that the global curvature of the 
interfaces is identically null as in $d=1+1$.

Finally, the height distributions rescaled according to Eq.~(\ref{eqansatz2}) 
are shown in Fig.~\ref{fig6}(b), where the excellent data collapse gives another 
evidence for their universality. Figure \ref{fig6}(b) also shows the generalized 
Gumbel distribution with parameter $m=9.5$ (with $S=0.332$ and $K=0.221$ and 
rescaled to mean $-2.33$ and variance $0.34$ \cite{noteGumbel}), which is a good fit of $\chi$ for 
the curved KPZ subclass in $d=2+1$~\cite{Oliveira13}. Moreover, rescaled 
distributions for $\omega=0$ are also shown in Fig. \ref{fig6}(b), which clearly 
show the existence of two different universal distributions for the underlying 
fluctuations of fixed-size and enlarging substrate KPZ subclasses in $d=2+1$. 
Again, the generalized Gumbel distribution, with parameter $m=6.0$ (with 
$S=0.424$ and $K=0.359$ and rescaled to mean $-0.90$ and variance $0.24$~\cite{noteGumbel}), 
provides a good fit of $\chi$ for almost five decades around the peak, as also 
observed in \cite{Oliveira13}.

\section{Spatial covariance}
\label{seccovspatial}

\begin{figure}[!t]
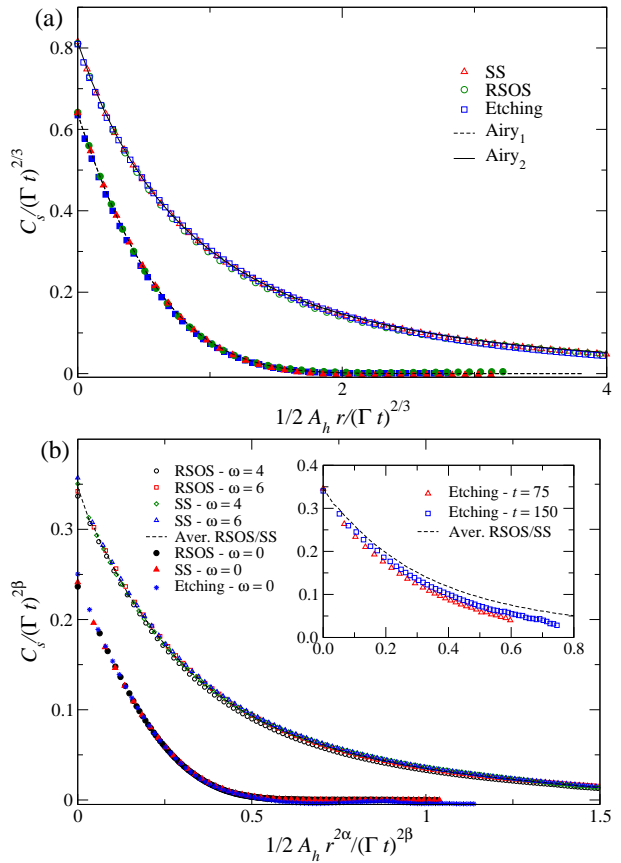

\includegraphics[width=8cm]{Fig7a.eps}
\includegraphics[width=8cm]{Fig7b.eps}
\caption{(Color online) (a) Rescaled spatial covariances for all investigated 
models in $d=1+1$ with $\omega=0$ (full) and $\omega=20$ (open symbols), for 
times $t=4000$ and 10000, respectively. (b) Rescaled spatial covariances in 
$d=2+1$ dimensions, for times $t=600$ ($\omega=0$), $t=250$ ($\omega=4$) and 
$t=150$ ($\omega=6$). Results for the Etching model on enlarging substrates are 
shown in the inset at two different times. Exponents $\alpha=0.395$ and $\beta=0.237$ were 
used for the rescaling.}
\label{fig7}
\end{figure}

Beyond the asymptotic height distribution, the limiting processes that describe 
the spatial profile of the flat and curved KPZ-class interfaces are exactly 
known in 1+1 dimensions and called the Airy$_1$ and Airy$_2$ processes, 
respectively~\cite{PraSpo3,*Sasa2005,*Borodin,Kriecherbauer.Krug-JPA2010}. We 
calculate the spatial covariance
\begin{equation}
 C_s(r,t) = \left\langle \tilde{h}(x,t) \tilde{h}(x+r,t) \right\rangle \simeq (\Gamma
t)^{2 \beta} \Psi[A_h r/(\Gamma t)^{2 \beta}],
\end{equation}
where $\tilde{h} \equiv h- \left\langle h \right\rangle$, $\Psi$ is a scaling 
function and $A_h = A$ in $1+1$ and $A_h = 0.6460 A$ in $2+1$ 
dimensions~\cite{healy2014}. Figure~\ref{fig7}(a) shows the rescaled spatial 
covariance for $d=1+1$ along with the Airy$_1$ and Airy$_2$ covariances. We find 
that the results for $\omega > 0$ and $\omega = 0$ are in good agreement with 
the Airy$_2$ and Airy$_1$ covariances, respectively, showing that the 
equivalence between expanding substrate systems and curved interfaces also holds 
for the spatial correlation.

In $d=2+1$, the spatial covariance for flat interfaces was numerically calculated only very recently~\cite{healy2014} and it was shown to be universal, as is the case for $d=1+1$.
However, there were no reports on the spatial covariance for curved interfaces up to the present work. Here, we determined the spatial covariance for the investigated models for both $\omega=0$ and $\omega>0$. The rescaled curves are presented in Fig. \ref{fig7}(b), where two universal curves for fixed-size and enlarging substrates are observed. These can be regarded as the (2+1)-dimensional analogue of the Airy$_1$ and Airy$_2$ covariances, respectively. For the Etching model on enlarging substrates, the rescaled curves do not converge yet within the examined time window, but are still approaching the asymptotic curve obtained for the RSOS and SS models (see inset of Fig. \ref{fig7}).

\section{Temporal covariance}
\label{seccovtemp}

Similarly to the spatial case, we can define the temporal covariance
\begin{equation}
 C_t(t,t_0) = \left\langle \tilde{h}(x,t_0) \tilde{h}(x,t) \right\rangle,
\end{equation}
which is expected to scale as $C_t(t,t_0) \simeq (\Gamma^2 t_0 t)^{\beta} \Phi(t/t_0)$~\cite{Kallabis99,Singha-JSM2005,TakeuchiJSP}. Similarly to the results for the spatial correlation, it is reasonable to expect that the scaling function $\Phi(x)$ is universal within each subclass and dimensionality. However, unlike the spatial correlation, no exact results have been obtained for the temporal correlation, even for $d=1+1$. Therefore, it is very important to check their universality by empirical approaches, such as simulations and experiments.

\begin{figure}[!t]
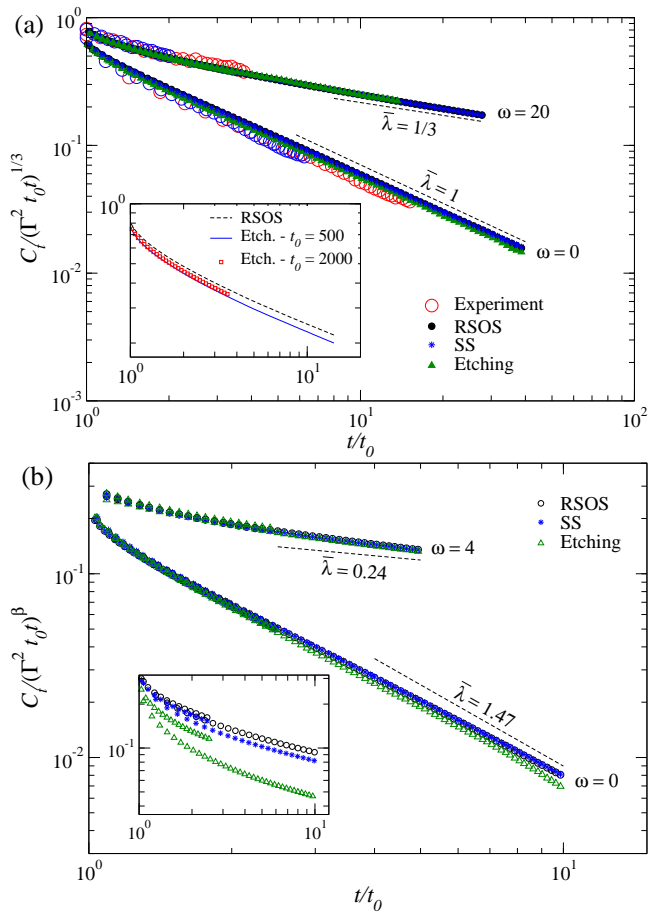

\includegraphics[width=8.5cm]{Fig8a.eps}
\includegraphics[width=8.5cm]{Fig8b.eps}
\caption{(Color online) (a) Rescaled temporal covariance for all investigated models (full symbols) in $d=1+1$ and the liquid-crystal experiment (open symbols) with flat and curved interfaces. In both cases, different initial times $t_0$ in the range $[100,2000]$ (models) and $[4s,25s]$ (experiment) were used. For Etching model and curved experimental surfaces, the curves are extrapolations. The raw data for Etching model are shown in the inset. (b) Rescaled temporal covariances for all models in $d=2+1$ dimensions and different initial times. For $\omega=4$, all curves in the main panel are extrapolations, while the non-extrapolated ones are shown in the inset.}
\label{fig8}
\end{figure}

Figure~\ref{fig8}(a) shows the rescaled temporal covariance for all investigated models in $d=1+1$ on fixed-size and enlarging substrates, along with the experimental data for flat and curved interfaces, respectively, generated in the electroconvection of nematic liquid crystals \cite{TakeuchiPRL,*TakeuchiSP,TakeuchiJSP} (already reported in \cite{TakeuchiJSP}). For fixed-size substrates, a good collapse of data for all models and initial times $t_0$ is observed. Moreover, a good agreement between the numerical and experimental covariances is also found. For enlarging substrates, the temporal covariances for the RSOS and SS models already reach an asymptotic function, clearly different from that for the fixed-size case, while finite-time effect seems to be more severe for the Etching model, similarly to the spatial covariance as reported in the inset of Fig.~\ref{fig7}(b). Similar effect was also observed for the curved interfaces in the liquid-crystal experiment (see the inset of Fig. 11(b) in Ref. \cite{TakeuchiJSP}).

To substantiate that the deviation from the asymptotic form is indeed a 
finite-time effect, we attempt an extrapolation of the finite-time data as 
follows. Assuming that $\eta$ is stochastic (as shown by past studies 
\cite{Oliveira12,Alves13,TakeuchiJSP,Ferrari.Frings-JSP2011}) and neglecting 
$\zeta$ fluctuations (as discussed above), we expect from the KPZ ansatz [Eq.~(\ref{eqansatz2})] that the leading correction in $\Phi$ is in the order of $t^{-\beta}$.
Using this expression and data at two different $t_0$ with fixed $t/t_0$, we can 
extrapolate to obtain the asymptotic temporal covariance $\Phi$. The data for 
the Etching model in the main panel of Fig.~\ref{fig8}(a) are obtained in this 
way from the raw data in the inset, and an excellent agreement with the raw data 
for the RSOS and SS models is achieved. For the experimental data of the curved 
interfaces, the same quality of the collapse and agreement is obtained, by 
assuming that the leading finite-time correction to $\Phi$ is 
$\mathcal{O}(t_0^{-2\beta})$. This may be related to the vanishing finite-time 
correction for the one-point second-order cumulant $\expct{\chi^2}_c$ for the 
curved case \cite{TakeuchiJSP}, but such relationship needs to be clarified by 
further analytical and empirical studies.

In $d=2+1$, for fixed-size substrates, we again find a good collapse of rescaled 
(non-extrapolated) temporal covariances for all models and initial times, as 
shown in Fig. \ref{fig8}(b). However, for enlarging systems the finite-time 
effects are larger and, even for the RSOS and SS models we do not attain a data 
collapse, as shown in the inset of Fig. \ref{fig8}(b). In these models, 
extrapolations for the correction $\mathcal{O}(t_0^{-\beta})$ fail to collapse 
the data, whereas a correction $\mathcal{O}(t_0^{-2 \beta})$ provides a nice 
collapse for both models and for different times used for the extrapolation 
(Fig.~\ref{fig8}(b)). For the Etching model, a good collapse of data is not 
achieved with corrections $\mathcal{O}(t_0^{-\beta})$ or $\mathcal{O}(t_0^{-2 
\beta})$. Instead, an apparent collapse is achieved 
with an intermediate exponent $\mathcal{O}(t_0^{-0.32})$, possibly arising from 
a mixture of both terms above, within the time window we investigated.

The strengths of the two correction terms, $\mathcal{O}(t_0^{-\beta})$ and 
$\mathcal{O}(t_0^{-2 \beta})$, may be related to the finite-time correction to 
the second-order cumulant $\expct{\chi^2}_c$. Indeed, the Etching model, whose 
correction to the second-order cumulant is known to be large \cite{FabioSS}, 
also exhibits the large finite-time corrections to the temporal covariance, as 
presented above. Although these two corrections are formally different as they 
concern equal-time and two-time properties, respectively, better understanding 
of such relationship will certainly serve for a more unambiguous determination 
of the universal functional forms for the temporal covariance.

In any case, our results in Fig.~\ref{fig8} show that the temporal covariance is 
clearly different between the fixed-size and enlarging systems, and that they 
agree, in $d=1+1$ dimensions, with the results for the flat and curved 
interfaces in the liquid-crystal experiment, respectively. The rescaled 
covariance $\Phi(t/t_0)$ converges to a universal function in each case and each 
dimensionality, as substantiated by the three models investigated here. 
Importantly, Kallabis and Krug \cite{Kallabis99} had conjectured that for long 
times $\Phi(x) \sim x^{-\bar{\lambda}}$ with $\bar{\lambda} = \beta + d_s/z$ for 
the flat interfaces, while $\bar{\lambda} = \beta$ was later proposed for the 
curved interfaces \cite{Singha-JSM2005}. Besides confirming these scaling 
relations in $d = 1+1$, our results suggest that they also seem to be valid for 
$d = 2+1$ (dashed lines in Fig.~\ref{fig8}(b)), though clear power laws are not 
yet reached within the time studied.

\section{Crossover from fixed-size to enlarging substrates}
\label{seccross}

Our enlarging-substrate systems are also convenient to study crossover between 
the fixed-size/flat and enlarging/curved subclasses, or, for $1+1$ dimensions, 
between the GOE and GUE TW distributions. Such inter-subclass crossover has also 
attracted great interest \cite{Corwin-RMTA2012}, both theoretically and 
experimentally. Analytical studies have mostly dealt with crossover in space for 
$1+1$ dimensions \cite{Corwin-RMTA2012}: for example, Borodin et al. 
\cite{Borodin.etal-CPAM2008} and Le Doussal \cite{ledoussal} considered an 
initial condition composed of a flat substrate for $x<0$ and a wedge (curved) 
one for $x>0$, and formulated crossover from the GOE to GUE TW distributions, or 
from the Airy$_1$ to Airy$_2$ processes, which takes place as one moves from $x 
\to -\infty$ to $x \to \infty$. In contrast, crossover in time remains out of 
the reach of analytical studies, as it requires understanding of the temporal 
covariance, but it has been recently addressed numerically and experimentally, 
for the crossover from the flat to stationary 
subclasses~\cite{TakeuchiCross,HealyCross,healy13}. Here, we investigate 
temporal crossover from the fixed-size/flat to the enlarging/curved subclasses 
(from GOE to GUE TW for $d=1+1$), by starting with an initial substrate such 
that $L_0 \gg \omega$. As the mean substrate size grows as $\expct{L} = L_0 + 
\omega t$, the characteristic crossover time is given by $t^{*} \sim 
L_0/\omega$. We therefore expect that, for $t \ll t^*$ (or equivalently $\omega 
\rightarrow 0$ or $L_0 \rightarrow \infty$), the system essentially behaves as a 
fixed-size system, while for $t \gg t^*$ the statistical properties of the 
enlarging/curved systems should take over.

\begin{figure}[!t]
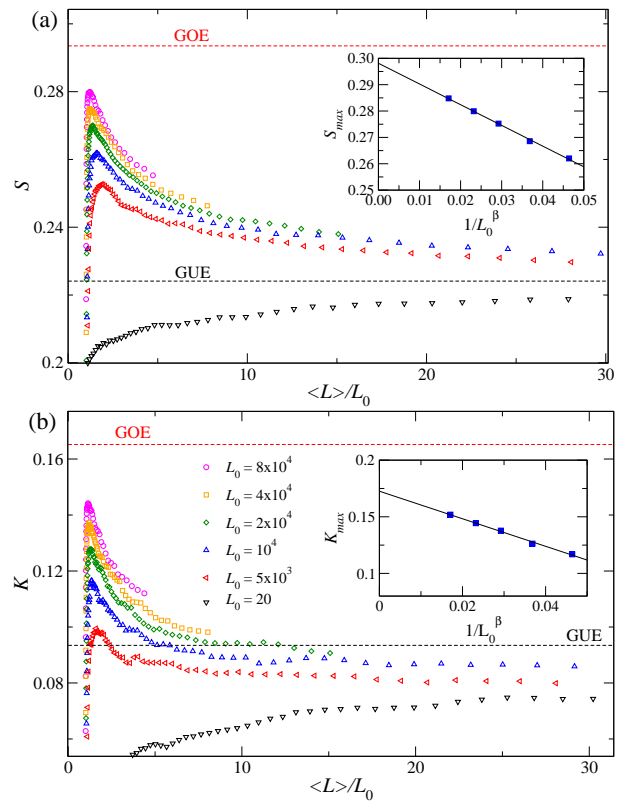

\includegraphics[width=8cm]{Fig9a.eps}
\includegraphics[width=8cm]{Fig9b.eps}
\caption{(Color online) Evolution of (a) the absolute value of the skewness and 
(b) the kurtosis for the one-dimensional RSOS model with $\omega=20$ and 
different initial system sizes. Insets show the maximal values against 
$L_{0}^{-\beta}$, with linear extrapolations to the asymptotic values.}
\label{fig9}
\end{figure}

This scenario is indeed consistent with our results shown in Figs. \ref{fig9}(a) 
and \ref{fig9}(b), where the skewness $S$ and kurtosis $K$ of the 
one-dimensional RSOS model are plotted as functions of $\expct{L}/L_0 \simeq  
t/t^{*}$. The cumulant ratios reach maxima near $t = t^*$, at some values close 
to those for the GOE TW distribution, and then approach the GUE TW values. As 
expected, the larger $L_0$ becomes, the closer the maximal values of the 
cumulant ratios are to the GOE TW values. Interestingly, these maxima $S_{\rm 
max}(L_0)$ and $K_{\rm max}(L_0)$ are found to vary linearly with $L_0^{-\beta}$ 
(insets). This allows us to extrapolate the asymptotic values $S_{\rm 
max}(\infty)=0.297(4)$ and $K_{\rm max}(\infty)=0.17(1)$, which agree with the 
GOE TW distribution ($S = 0.2935, K=0.1652$). Similar results were also obtained 
for all models in $2+1$ dimensions (data not shown).

\section{Conclusions}
\label{secconcl}

In this article, we studied typical models of the KPZ class, on flat substrates 
enlarging at a constant rate $\omega$. While the growth exponent $\beta$ is the 
same for fixed-size ($\omega=0$) and enlarging ($\omega > 0$) substrates, the 
height distribution does change: for the fixed-size case, it is given by the 
universal distribution for the flat interfaces (GOE TW in $d=1+1$), while for 
the enlarging case the distribution for the curved interfaces arises (GUE TW in 
$d=1+1$). We also reached the same conclusion for the spatial and temporal 
covariances. In particular, we found the Airy$_1$ and Airy$_2$ covariance for 
the spatial correlation of the fixed-size and enlarging systems in $d=1+1$, 
respectively, as well as agreement with the Kallabis-Krug conjecture on the 
temporal covariance \cite{Kallabis99,Singha-JSM2005,TakeuchiJSP}. Moreover, we 
also studied $(2+1)$-dimensional systems and found clear agreement in the height 
distribution, with the functional forms previously obtained numerically for the 
curved and flat interfaces.

All these results indicate that the interfaces growing on enlarging substrates 
share the same statistical properties as the curved interfaces, despite the fact 
that the global curvature in these enlarging systems is kept exactly null. This 
suggests that the substrate enlargement is possibly more relevant for the 
realization of the ``curved interface'' subclass than the global curvature 
itself. Indeed, to our knowledge, all interfaces deemed ``curved'' in previous 
work (e.g., 
\cite{Kriecherbauer.Krug-JPA2010,Corwin-RMTA2012,Johansson-CMP2000,PraSpo1, 
Tracy.Widom-CMP2009,TakeuchiSP,TakeuchiJSP,SasaSpo1,Oliveira13}) evolve within a 
zone of activity that grows linearly in time. The activity zone  
corresponds to the growing circumference for the usual circular interfaces, but 
this concept is also valid for the ASEP with the step initial 
condition~\cite{Johansson-CMP2000,Tracy.Widom-CMP2009}, in which particles can 
move only within a linearly growing area around the origin. We hope that the 
relevance of such substrate enlargement to the ``curved interface'' subclass 
will be further investigated on a mathematical or theoretical basis; for this, 
the so-called characteristic lines 
\cite{Ferrari-JSM2008,Corwin.etal-AIHPBPS2012} may be a useful concept, which 
describe the directions of the fluctuation propagation in space-time.

Beyond those asymptotic universal quantities, finite-time behavior was also 
characterized. We found a logarithmic correction in the height evolution 
[Eq.~(\ref{eqansatz2})] when the substrate is enlarging. We consider that this 
correction is a consequence of column duplications adopted in our time evolution 
rule for enlarging substrates. Furthermore, crossover from the fixed-size (flat) 
to the enlarging substrate (curved) subclasses (GOE to GUE TW distributions in 
$d=1+1$), which takes place in the course of time evolution, has also been 
characterized.

As a final remark, we stress that our simulation method based on the substrate 
enlargement provides a powerful tool to study statistical properties of the curved 
interface subclass, since this produces isotropic interfaces on a lattice. This is not 
the case of the usual growth models on lattice, such as the Eden model, which is 
known to produce an anisotropic interface even from a single point seed 
\cite{Stauffer86,Paiva07,Alves13} reflecting the lattice structure of the model. 
Having access to isotropic interfaces instead is essential to study statistical 
properties of interest numerically, because then we can use all spatial points 
to improve statistics and to define the spatial correlation functions 
unambiguously. Indeed, in this article, this allowed us to determine the 
two-point spatial correlation function for the enlarging/curved systems in $2+1$ 
dimensions, for the first time, numerically. Our method of enlarging substrates 
therefore provides a useful platform to study statistical properties of the KPZ 
class in higher dimensions, and of other universality classes for fluctuating 
surface growth problems.

\acknowledgments

The authors thank R. Cuerno and I. Corwin for helpful discussions. This work is 
supported in part by CNPq, CAPES and FAPEMIG (Brazilian agencies) and by KAKENHI 
(No. 25707033 from JSPS and No. 25103004 ``Fluctuation \& Structure'' from MEXT, 
in Japan).

\bibliography{KPZGrowing}

\begin{thebibliography}{10}
\bibitem{noteGumbel}%
  \BibitemOpen
  \bibinfo {note} {Although the generalized Gumbel distributions with $m = 6.0$
  and $9.5$ serve as good fits to the distribution for the flat and curved KPZ
  subclasses, respectively, in $d=2+1$ dimensions \cite{Oliveira13}, no
  theoretical argument suggests that they are the true asymptotic
  distributions. Therefore, the values of their skewness and kurtosis are not
  exactly those for the KPZ class. For the KPZ class values, one should refer
  to numerical estimates, reported in \cite{healy12,healy13,Oliveira13}}%
  \bibAnnoteFile{NoStop}{noteGumbel}%
\end{thebibliography}

\end{document}